\documentclass[twocolumn,prl,showpacs,aps,superscriptaddress]{revtex4-1}

\usepackage[english]{babel}
\usepackage[ansinew]{inputenc}
\usepackage{times}
\usepackage{graphicx}
\usepackage{graphics}
\usepackage{amsmath}
\usepackage{amsfonts}
\usepackage{amssymb}
\usepackage{amsbsy}
\usepackage{dsfont}
\usepackage{epstopdf}
\usepackage{makeidx}
\usepackage{subfigure}
\usepackage{color} 
\usepackage{pgf}
\usepackage{bm}
\usepackage{tikz} \usepackage[normalem]{ulem}

\newcommand{\be}{\begin{equation}}
\newcommand{\ee}{\end{equation}}
\newcommand{\bea}{\begin{eqnarray}}
\newcommand{\eea}{\end{eqnarray}}

\makeindex
\begin{document}

\title{Topological phase transitions in the photonic spin Hall effect}

\author{W. J. M. Kort-Kamp}
\affiliation{Center for Nonlinear Studies and Theoretical Division, Los Alamos National Laboratory, MS B258, Los Alamos, NM 87545, USA}
\email{kortkamp@lanl.gov}
\date{\today}

\begin{abstract}
The recent synthesis of  two-dimensional staggered materials opens up burgeoning opportunities
to study optical spin-orbit interactions in semiconducting Dirac-like systems. We unveil topological phase transitions in the photonic spin Hall effect in the graphene family materials. It is shown that an external static electric field and a high frequency circularly polarized laser allow for active on-demand manipulation of electromagnetic beam shifts.  The spin Hall effect of light presents a rich dependence with radiation degrees of freedom, material properties, and features non-trivial topological properties. We discover that photonic Hall shifts are sensitive to spin and valley properties of the charge carries, providing a unprecedented pathway to investigate spintronics and valleytronics in staggered 2D semiconductors. 
\end{abstract}
%
\maketitle
At macroscopic scales electromagnetic radiation's spatial and polarization degrees of freedom are independent quantities that can be accurately described by traditional geometric optics. A different landscape takes place in the subwavelength regime where emergent photonic spin-orbit interactions (SOI) culminate in spin-dependent changes in light's spatial properties \cite{Bliokh2015, Bliokh20152}. A striking optical phenomena originating from SOI is the spin Hall effect of light (SHEL), which corresponds to the shift of photons with contrary chirality to opposite sides of a finite beam undergoing reflection/refraction \cite{Onoda2004, Bliokh2006, Hosten2008, Ling2017}. The SHEL is universal to any interface and represents a remarkable failure of Fresnel's and Snell's formulas at the nanoscale. It exhibits a unique potential for applications in precision metrology, including bio-sensing \cite{Yin2006}, nanoprobing \cite{Herrera2010},  and thin films and multilayer graphene characterization  \cite{Zhou20122, Qiu2014, Zhou2012}. It has also been used to 
identify different absorption mechanisms in bulk semiconductors \cite{Menard2009,Menard2010}. 

Staggered two-dimensional semiconductors \cite{Gomez2016, Manix2017, Molle2017}, including silicene \cite{PhysRevLett.108.155501}, germanene \cite{Davila-2014}, and stanene \cite{RIS_0,Saxena2016} are monolayer materials made of Silicon, Germanium, and Tin atoms, respectively, arranged in a honeycomb lattice. Unlike
graphene \cite{RevModPhys.81.109},
these materials are nonplanar and possess intrinsic spin-orbit coupling that results in the opening of a gap in their electronic band structure. Under the influence of external static and circularly polarized electromagnetic fields the four Dirac gaps are in general nondegenerate and the monolayer may be driven through several phase transitions involving topologically non-trivial states \cite{PhysRevB.86.195405, PhysRevLett.110.197402, Ezawa2013, Ezawa-2015, Lopez2017}.  Previous studies on SHEL in the graphene family have been restricted to graphene \cite{Grosche2015, KortKamp2016, Cai2017}, therefore overlooking the role of finite staggering, 
spin-orbit coupling, and spin/valley dynamics.
The interplay between topological matter and SHEL was
considered in magnetic field biased bulk materials with axion coupling \cite{Wen2013}. In this letter we take advantage of the crossroads between topology, phase transitions, spin-orbit interactions, and Dirac physics in staggered 2D semiconductors to uncover magnetic field free topological phase transitions in the photonic spin Hall effect. We show that the SHEL depends on the topological invariant describing each phase  and it allows to probe the spin and valley properties of charge carriers throughout different phase transitions. The marriage of spinoptics, spintronics, and valleytronics in 2D 
semiconductors opens a promising route to investigate emergent electronic and photonic phenomena in the graphene family.

\begin{figure}
\centering
\includegraphics[width=\linewidth]{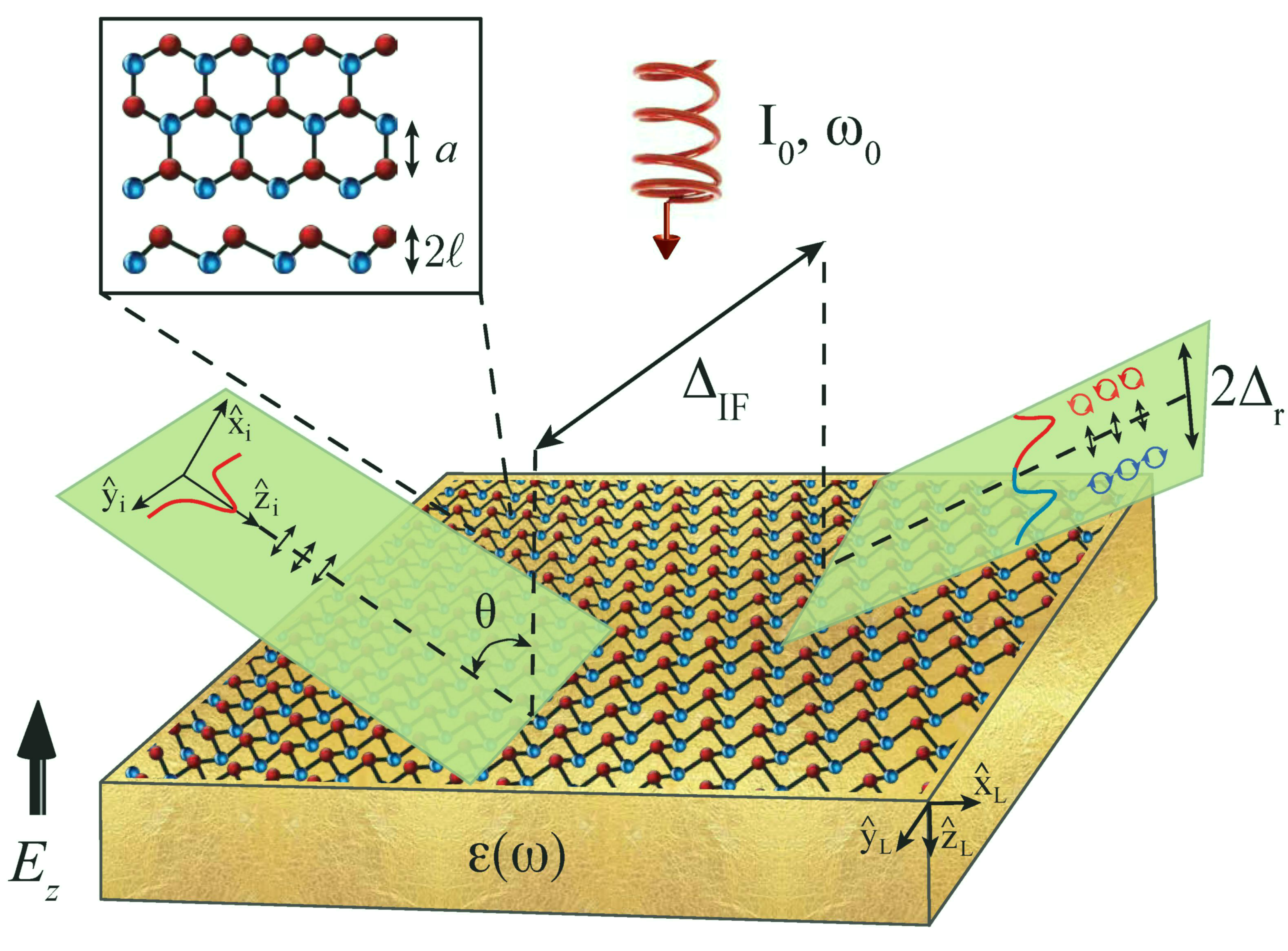}
\caption{Schematic representation of the system under study. The inset exhibits  the top and side views of staggered graphene family materials. Lattice constant and staggering length values are $a = (3.86, 4.02, 4.7)\ \text{\AA}$ and $\ell = (0.23, 0.33, 0.4)\ \text{\AA}$  for silicene, germanene, and stanene, respectively \cite{Ezawa-2015}.}
\label{Fig1}
\end{figure}

Let us consider that a Gaussian beam \cite{BornWolf} of frequency $\omega$ impinges at an angle $\theta$ on a staggered monolayer placed on top of a substrate of dielectric constant $\varepsilon$. The incident beam ${\bf E}_{i}= \mathcal{A}(y_i,z_i) [f_p \hat{{\bf x}}_i + f_s \hat{{\bf y}}_i - i f_s  k y_i (\Phi+ik z_i)^{-1} \hat{{\bf z}}_i] $ is confined in the $y$-direction only. Here, $\mathcal{A}(y,z)= [2/\pi w_0^2 (1+k^2 z^2/\Phi^2)]^{1/4} e^{i k z - k^2y^2/2(\Phi+ik z)}$ is the Gaussian amplitude, $w_0$ is the beam waist, $k=\omega/c$ is the wavenumber, and $\Phi=k^2w_0^2/2$ is the Rayleigh range. The polarization of the beam is given by the complex unit vector $\hat{{\bf f}}= f_{p}{\bf\hat{x}}_{i}+f_{s}{\bf\hat{y}}_{i}$, where $f_p = 1, f_s =0$ ($f_p=0, f_s=1$) corresponds to a linearly polarized transverse magnetic (electric) state. Relevant unit vectors are defined in Fig. \ref{Fig1}.  In addition to the Gaussian beam, the system is subject to a static electric field $E_z$ and to a circularly polarized plane wave of intensity $I_0$  and frequency $\omega_0 \gg \omega$ propagating along $z$-direction. Whilst $E_z$ generates an electrostatic potential $2\ell E_z$ between the two inequivalent sub-lattices of the monolayer (see inset in Fig. \ref{Fig1}) \cite{PhysRevB.86.195405, PhysRevLett.110.197402}, the high frequency laser modifies the electronic band structure of the material and chiral states may arise even in the absence of magnetic fields \cite{Ezawa2013, Ezawa-2015}.   As a consequence, the material presents a non-zero Hall conductivity that induces polarization conversion of the incident radiation. 

We shall restrict our discussion to the SHEL in reflection since similar qualitative results hold for the transmitted wave.  Analytical results for  the reflected electromagnetic field ${\bf E}_r$ can be derived within the paraxial approximation. Expanding the incident beam in a plane wave basis and enforcing standard boundary conditions for each component, one obtains
\begin{eqnarray}
\!\!\!\!\!\! {\bf E}_r  \simeq \mathcal{E}_r   \!\!\!  \!\!\!  & &  \!\!  \left[(1\!+\!i\rho_R)\mathcal{A}(y_r\!-\!\tilde{\delta}_1,z_r) - \rho_I\mathcal{A}(y_r\!-\!\tilde{\delta}_2,z_r) \right] \hat{{\bf e}}_{-}  \cr
 &+& \!\!  \left[(1\!-\!i\rho_R)\mathcal{A}(y_r\!-\!\tilde{\delta}_2,z_r) + \rho_I\mathcal{A}(y_r\!-\!\tilde{\delta}_1,z_r) \right] \hat{{\bf e}}_{+}  ,
\label{FieldSHEL}
\end{eqnarray}
where $\mathcal{E}_r$ is a constant, $\rho = \rho_R + i\rho_I = (f_s r_{\!_{ss}} + f_p r_{\!_{sp}}) / (f_p r_{\!_{pp}} + f_s r_{\!_{ps}})$, and $r_{\!_{ij}}$ are the Fresnel's reflection coefficients for incoming $j$- and outgoing $i$-polarized plane waves. Expressions for $r_{\!_{ij}}$ in terms of the monolayer's conductivity can be derived  by modeling the 2D material as a surface density current at $z = 0$ \cite{KortKamp2015}.  Besides, 
$\hat{{\bf e}}_{\pm} = [\hat{{\bf x}}_r\pm i(\hat{{\bf y}}_r-\beta y_r\hat{{\bf z}}_r)]/\sqrt{2}$ are left $(+)$ and right $(-)$ circularly polarized unit vectors with $\beta = ik/(\Phi+ikz_r)$, and ${\bf\hat{x}}_r={\bf\hat{x}}_i - 2 {\bf\hat{x}}_L ({\bf\hat{x}}_i \cdot {\bf\hat{x}}_L)$, 
${\bf\hat{y}}_r={\bf\hat{y}}_i$, and
${\bf\hat{z}}_r={\bf\hat{z}}_i - 2 {\bf\hat{z}}_L ({\bf\hat{z}}_i \cdot {\bf\hat{z}}_L)$. Note that the reflected beam corresponds to the superposition of left and right circularly polarized states given by the  sum of two Gaussians centered at $\tilde{\delta}_1$ and $\tilde{\delta}_2$ and weighted by the real and imaginary parts of $\rho$. For low-dissipative materials ($\rho_I \ll \rho_R$), each component of the field reduces to single Gaussians with photons of opposite helicity  shifted by $\tilde{\delta}_1$ (right) and $\tilde{\delta}_2$ (left). The complex displacements $\tilde{\delta}_l $ can be conveniently written as $\tilde{\delta}_l = \tilde{\Delta}_{\textrm{IF}} + (-1)^l \tilde{\Delta}_{\textrm{r}}$ $(l = 1, 2)$, where $\tilde{\Delta}_{\textrm{IF}} = Y_p (|\rho|^2+ 1)^{^{-1}} ( f_{p}+f_{s}r_{\!_{ps}}/r_{\!_{pp}})^{^{-1}}+ \{ p\leftrightarrow s \}$,  $Y_p = i  f_s (r_{\!_{pp}}+ r_{\!_{ss}}) \cot \theta/ k r_{\!_{pp}}$, $Y_s = -Y_p|_{p \leftrightarrow s}$, and $\tilde{\Delta}_{\textrm{r}}$ is obtained from $\tilde{\Delta}_{\textrm{IF}}$ by replacing $Y_p \rightarrow i\rho Y_p$ and $Y_s \rightarrow -iY_s/\rho^{*}$. We mention that $\Delta_{\textrm{IF}} = \textrm{Re}[\tilde{\Delta}_{\textrm{IF}}] $ and $\Theta_{\textrm{IF}} = k \textrm{Im}[\tilde{\Delta}_{\textrm{IF}}]/\Phi $ are the spatial and angular Imbert-Fedorov shifts \cite{Bliokh2006, Fedorov1955, Imbert1972}, respectively. The SHEL shifts of the intensity distribution centroid for left and right circular polarizations can be cast as
\begin{equation}
{\Delta}^{\pm}_{\textrm{SHEL}}\! =\! \Delta_{\textrm{IF}}  \pm \Delta_{\textrm{r}} \! = \! \Delta_{\textrm{IF}} \pm  \textrm{Re}\!\left[\!\dfrac{\tilde{\Delta}_{\textrm{r}}(1+\rho^{*2})+ 2\rho_I \tilde{\Delta}_{\textrm{IF}}}{1+|\rho|^2}\!\right] .
\label{SHEL_Shift}
\end{equation}

In order to evaluate ${\Delta}^{\pm}_{\textrm{SHEL}} $ one needs the optical conductivity tensor of the monolayer at temperature $T$ and doping $\mu$. Using Kubo's formalism \cite{Kubo-I-1957, Maldague} one obtain $\sigma_{ij}(\omega,\mu,T)=\sum_{\eta,s} \int_{-\infty}^{\infty}\tilde{\sigma}_{ij}^{\eta s}\!(\omega, E)/4 k_{\rm B}T\cosh^{2}\!\left[(E-\mu)/2 k_{\rm B}T\right] dE$, with zero temperature conductivities $\tilde{\sigma}_{ij}^{\eta s} (\omega, \mu)$ given by
\begin{eqnarray}
\dfrac{\tilde{\sigma}_{xx}^{\eta s}}{\sigma_0/2\pi}\! &=&\!
\dfrac{4\mu^2\!-\!{\Delta_{s}^{\eta}}^2}{2\hbar \mu\Omega}\Theta(2\mu\!-\!|\Delta_{s}^{\eta}| )\! +\!\! \left[\!1\! -\! \dfrac{{\Delta_{s}^{\eta}}^2}{\hbar^2\Omega^2}\!\right]\!\! {\rm{tan}}^{^{\!\!-1}}\!\!\!\left[\!\dfrac{\hbar\Omega}{M}\!\right]\!\! \cr
&+& \! \dfrac{{\Delta_{s}^{\eta}}^2}{\hbar\Omega M}, \ \ \ \
\dfrac{\tilde{\sigma}_{xy}^{\eta s}}{\sigma_0/2\pi}\!=\!
\frac{2\eta\Delta_{s}^{\eta}}{\hbar \Omega} \tan^{^{\!\!-1}}\!\!\!\left[\!\frac{\hbar\Omega}{M}\!\right]. 
\label{sigma_Dirac_Cone}
\end{eqnarray}
\begin{figure}
\centering
\includegraphics[width=1\linewidth]{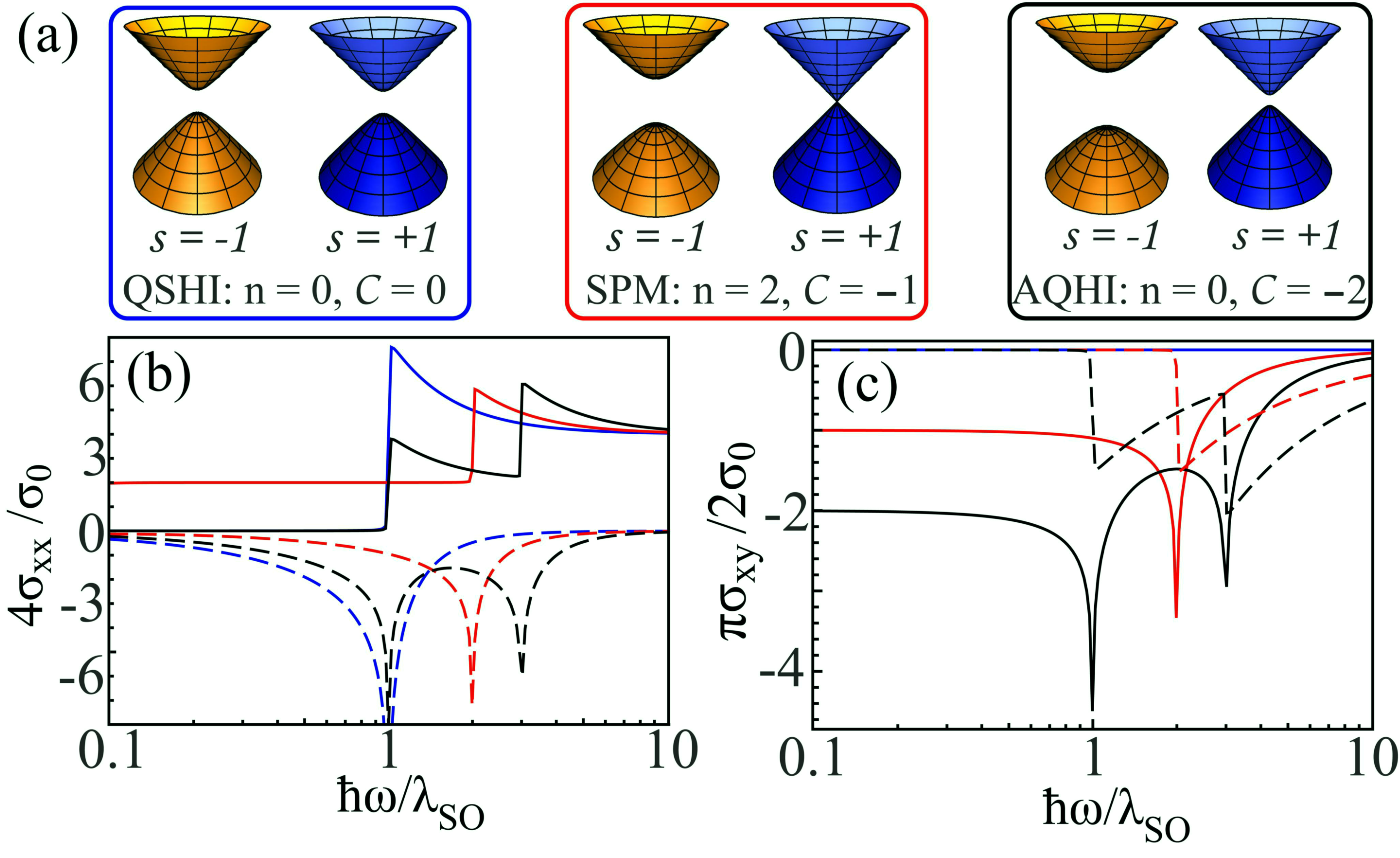}
\caption{(a) Electronic band structure of graphene family semiconductors for spin down ($s = -1$) and up ($s = +1$). $n$ and $\mathcal{C}$ are the number of closed gaps and Chern number, respectively. Parameters are $\{e\ell E_z/\lambda_{\rm SO}, \Lambda/\lambda_{\rm SO}\} = \{0, 0\}$ (left), $\{0,1\}$ (middle), and $\{0,2\}$ (right). Valleys $K (\eta = +1)$ and $K^{'} (\eta = -1)$ have similar behavior for $E_z = 0$. Real (solid) and imaginary (dashed) components of (b) $\sigma_{xx}$ and (c) $\sigma_{xy}$, 
versus $\omega$ for the phases described in (a). The layer is neutral, $k_{\rm B}T = 10^{-2}\lambda_{\rm SO}$, and $\hbar \Gamma = 0.002\lambda_{\rm SO}$.
}
\label{Fig2}
\end{figure}
%
Here, $\tilde{\sigma}_{yy}^{\eta s}=\tilde{\sigma}_{xx}^{\eta s}$, $\tilde{\sigma}_{yx}^{\eta s}=-\tilde{\sigma}_{xy}^{\eta s}$, $\sigma_0 = e^2/4\hbar$, $\Omega = -i\omega + \Gamma$, where $\Gamma$ is the scattering rate, and $M=\text{max}( |\Delta_{s}^{\eta}|,2 |\mu|)$. 
The mass gap  $\Delta_{s}^{\eta}  =  \eta s\lambda_{\text{SO}} - e  \ell E_{z} - \eta\Lambda$ depends on the strength of the intrinsic spin-orbit coupling $\lambda_{\rm SO}$ ($\sim$2, $\sim$20, and $\sim$300 meV for silicene, germanene, and stanene \cite{Gomez2016}) and the spin ($s = \pm 1$) and valley ($\eta = \pm 1$) numbers.
The coupling constant between the monolayer and the high frequency laser is $\Lambda =  \pm 8\pi\alpha v_F^2 I_0/\omega_0^3$, where  $\alpha$ is the fine structure constant, $v_F$ is Fermi velocity, and the $+ (-)$ sign corresponds to left (right) circular polarization. The external fields allow for extraordinary manipulation of the four Dirac gaps, therefore enabling active control of the optoelectronic response of the monolayer (see Fig. \ref{Fig2}). For instance, at $E_z = \Lambda = 0$ the system behaves as a quantum spin Hall insulator (QSHI). If we increase $\Lambda$ while keeping $E_z = 0$, the Dirac gaps for $s = +1$ decrease. At $\Lambda = \lambda_{\rm SO}$ these gaps close and the system undergoes a topological phase transition from the QSHI phase to the spin polarized metal (SPM) phase. Further increasing $\Lambda$ results in reopening the gaps and the system reaches the anomalous quantum Hall insulator (AQHI) phase. Similar transitions can be obtained by changing $E_z$. As a consequence, staggered semiconductors of the graphene family present a rich phase diagram \cite{Ezawa2013}. 
Many of the phases have non-trivial topological features which can be characterized by a topological invariant, namely the Chern number $\mathcal{C} = \sum_{\eta,s}^{'} \eta\textrm{sign}[\Delta^{\eta}_s]/2$, where the prime indicates that only open gaps should be summed over. For a left circularly polarized laser, $\mathcal{C} = 0, -1, -2$ in the QSHI, SPM, and AQHI phases, respectively (see Fig. \ref{Fig3} for other phases and their corresponding Chern number). In Fig. \ref{Fig2}b,c we illustrate the behavior of  $\sigma_{xx}$ and $\sigma_{xy}$ as a function of $\omega$ in the aforementioned phases. Note that ${\rm Re}[\sigma_{xy}]$ (${\rm Re}[\sigma_{xx}]$) is proportional to $\mathcal{C}$ ($n$)
at low frequencies.
\begin{figure}
\centering
\includegraphics[width=\linewidth]{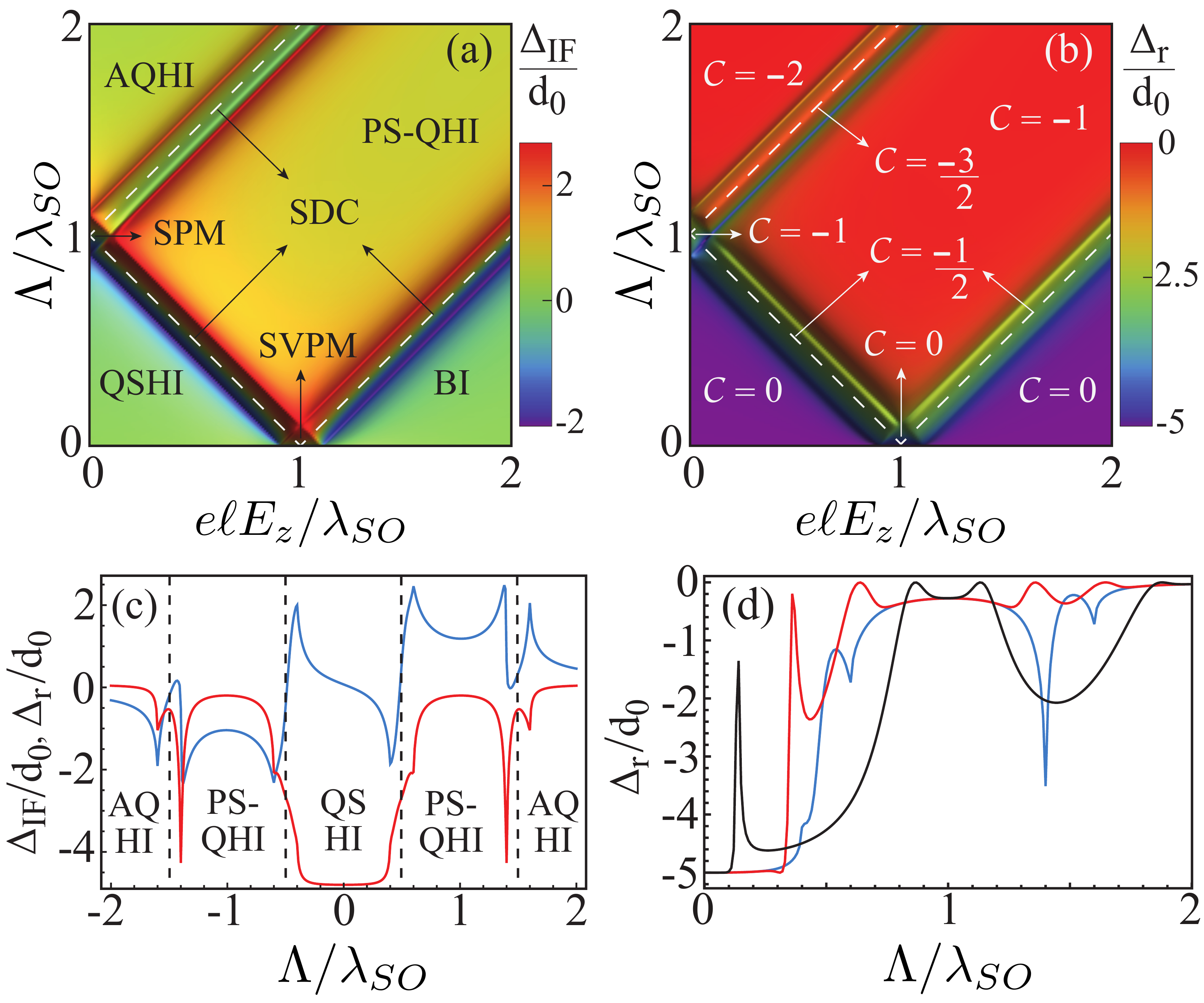}
\caption{Phase diagram of (a) $\Delta_{\rm IF}$ and (b) $\Delta_{\rm r}$  for suspended neutral staggered monolayers. The electronic phases are quantum spin Hall insulator (QSHI), spin-valley polarized metal (SVPM), band insulator (BI), single Dirac cone (SDC), spin-polarized metal (SPM), anomalous quantum hall insulator (AQHI), and polarized-spin quantum Hall insulator (PS-QHI). 
(c)  $\Delta_{\rm IF}$ (blue) and $\Delta_{\rm r}$ (red) as a function of $\Lambda$ for $\mu = 0$.
Dashed lines mark  phase transition boundaries. 
(d) $\Delta_{\rm r}$ versus $\Lambda$ for $\mu/\lambda_{\rm SO}= 0.05$ (blue), $0.1$ (red), and $0.2 $ (black). 
The 
beam is $s$-polarized, $\hbar \omega = 0.1\lambda_{SO}$, $\theta = \pi/4$, $d_0 = \hbar c/\lambda_{\rm SO}$, and $e\ell E_z/\lambda_{\rm SO} = 0.5$ in (c) and (d)
\cite{comment1}.
}
\label{Fig3}
\end{figure}
%

In Fig. \ref{Fig3} we unveil the role of topology and spin-orbit interactions in the photonic spin Hall effect for a suspended  staggered  monolayer. Panels (a) and (b) show the phase diagram for the Imbert-Fedorov and relative SHEL displacements, respectively. The dashed white lines 
mark semimetallic (SVPM, SPM, and SDC) phases where at least one gap is closed. Note that both shifts present signatures of the phase transitions taking place in the material. In the insulating QSHI, BI, AQHI, and PS-QHI phases all Dirac gaps are open with $\Delta_{\rm IF}$ and $\Delta_{\rm r}$ being weakly affected by the external fields.  However, close to phase boundaries a strong modulation of the photonic shifts is possible. For instance, $\Delta_{\rm IF}$ changes from positive to negative values near semimetallic phases even for small variations of $E_z$ or $\Lambda$ (see Fig. \ref{Fig3}c). We point out that $\Delta_{\rm r}$ ($\Delta_{\rm IF}$) gives the dominant contribution to $\Delta_{\rm SHEL}^{\pm}$ in QHSI and BI (AQHI and PS-QHI) phases. Both $\Delta_{\rm IF}$ and $\Delta_{\rm r}$ are equally relevant as we approach semimetallic states.   The width of the region around a phase transition where abrupt variations in the SHEL occur is determined by the frequency of the impinging beam. Indeed, for beam frequencies smaller than all Dirac gaps no electrons can be excited from the valence to the conduction band. On the other hand, for phase space regions where $|\Delta_s^{\eta}|< \hbar \omega$ the Gaussian wave may generate electron-hole pairs, modifying the layer's conductivity. 

A closer inspection of the regions next to phase transition boundaries shows that topology plays a crucial role in the photonic spin Hall effect. Let us consider that we drive the system through distinct phases by increasing $\Lambda$ while keeping $E_z$ fixed. For $e\ell E_z/\lambda_{\rm SO} = 0.5$ (Fig. \ref{Fig3}c) we note that $\Delta_{\rm IF}$ and $\Delta_{\rm r}$ present a smooth transition from the QSHI to the PS-QHI phase while crossing the SDC phase with $\mathcal{C}_{_{\rm SDC}} = \pm 1/2$. However,  a highly non-monotonic behavior appears near the $\mathcal{C}_{_{\rm SDC}} = \pm 3/2$ semimetallic phases as the system goes from the PS-QHI to the AQHI state. As a matter of fact, we checked that $\Delta_{\rm IF}$ and $\Delta_{\rm r}$ present a non-trivial (monotonous) dependence with $E_z$ and $\Lambda$ in any transition where all (at least one) phases involved have $\mathcal{C} \neq 0$ ($\mathcal{C} = 0$). Since the Chern number depends on the characteristics of the open (closed) gaps,
this result suggests that the SHEL is sensitive to the spin and valley numbers of the charge carries affected by the transition.  This is in remarkable contrast with effects of topological phase transitions in quantum fluctuations \cite{Lopez2017}, where electromagnetic interactions in semimetallic phases depend on the number of closed gaps but not on the values of $\eta$ and $s$ (thus, independent of ${\mathcal{C}}$). Hence, light beam shifts could be used to probe the dynamics of specific Dirac gaps across a phase transition. Finally, note that $\Delta_{\rm IF}$ 
changes sign whenever the polarization of the high frequency laser is swapped from left ($\Lambda > 0$) to right ($\Lambda < 0 $), enabling control of the SHEL direction. 
Contrariwise, $\Delta_{\rm r}$ is an even function of $\Lambda$. Inversion of $E_z$ direction does not affect the sign of any shifts (not shown). 

Figure \ref{Fig3}d depicts the impact of doping on the SHEL phase transitions for $e\ell E_z/\lambda_{\rm SO} = 0.5$. In the regions where $|\mu| < |\Delta^{\eta}_{\rm s}|$ the effect of doping can be neglected and the shifts correspond 
to those due to neutral monolayers. However, close to phase transition boundaries where $|\mu| > |\Delta^{\eta}_{\rm s}|$, intraband transitions take place and the influence of doping 
is appreciable. In contrast to neutral layers, for instance, a peak in $\Delta_{\rm r}$ emerges around $\Lambda/\lambda_{\rm SO} =  0.5$ for $\mu=0.05\lambda_{\rm SO}$. Further increasing the Fermi energy allows to shift the peak's position to lower values of $\Lambda$.  Similarly, near the phase transition at $\Lambda/\lambda_{\rm SO} =  1.5$ a crossover in $\Delta_{\rm r}$ from two narrow deeps to a broad one occurs as doping is increased.
\begin{figure}
\centering
\includegraphics[width=\linewidth]{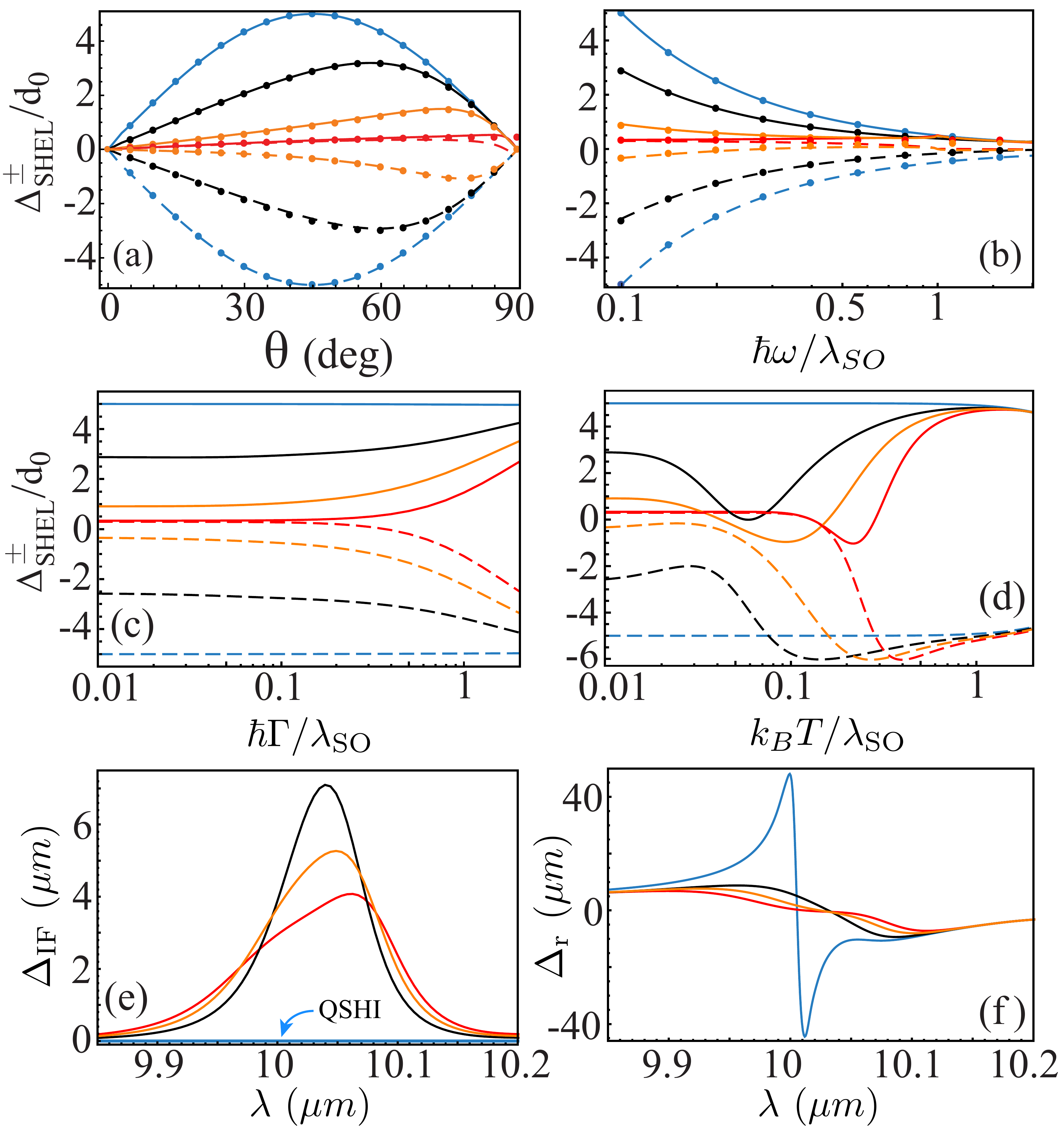}
\caption{(a)-(d) Photonic spin Hall shifts  $\Delta^{-}_{\rm SHEL}$ (solid) and $\Delta^{+}_{\rm SHEL}$ (dashed) as a function of different parameters
for a suspended neutral monolayer and $s$-polarization. Dotted curves correspond to the 
results in Eq. (\ref{Expansion_SHEL}). (e) $\Delta_{\rm IF}$ and (f) $\Delta_{\rm r}$ versus wavelength
for a $p$-polarized 
beam impinging on a stanene coated SiC substrate.
In all panels $\{e\ell E_z/\lambda_{\rm SO}, \Lambda/\lambda_{\rm SO} \}$ = $\{0,0\}$ (QSHI, blue), $\{0, 1\}$ (SPM, black), $\{0, 2\}$ (AQHI, red), and $\{0.5, 1.5 \}$ (SDC, orange). 
Other parameters are the same as in Fig. \ref{Fig3}. 
}
\label{Fig4}
\end{figure}

In order to get further insight on the interplay between topological phase transitions and the SHEL, we perform an expansion of Eq. (\ref{SHEL_Shift}) for small frequencies and dissipation for a neutral monolayer. For the sake of simplicity we assume that $\sigma_{ij}$ can be described by its zero temperature expression. For a $s$-polarized beam and up to leading order in $\omega$, $\Gamma$, one has
\begin{eqnarray}
&&\left.\Delta_{\rm SHEL}^{\pm ({\rm s})}\right|_{_{\substack{{\rm {\footnotesize QSHI,\ BI}}\\ {\rm {\footnotesize SVPM}}}}}^{\mathcal{C} = 0}\! =\! -c\sin\!\theta \cos\!\theta\! \left[\dfrac{\hbar}{|{\Delta}_e|}C_e\cos\!\theta \pm \dfrac{1}{\omega} \right] , \nonumber \\ \cr
&&\left.\Delta_{\rm SHEL}^{\pm ({\rm s})}\right|_{_{\substack{{\rm {\footnotesize SPM}}\\ {\rm {\footnotesize SDC}}}}}^{\mathcal{C} \neq 0}\! =\! \dfrac{-c \sin\!\theta\cos\!\theta}{n^2\pi^2+64\mathcal{C}^2\cos^2\!\theta}\! \left[\!\dfrac{\hbar}{|{\Delta_e}|}\!\dfrac{64\mathcal{C}}{3}\cos\!\theta \pm \! \dfrac{n^2\pi^2}{\omega}\! \right] ,\nonumber \\ \cr
&&\left.\Delta_{\rm SHEL}^{\pm ({\rm s})}\right|_{_{\substack{{\rm {\footnotesize AQHI}}\\ {\rm {\footnotesize PS-QHI}}}}}^{\mathcal{C} \neq 0}\!=\! -\dfrac{\hbar c}{3|{\Delta_e}|}\sin\!\theta\! \left[\dfrac{1}{\mathcal{C}} \pm \dfrac{Z_0\sigma_0\Gamma}{\pi \omega} \right] ,
\label{Expansion_SHEL}
\end{eqnarray}
where $C_e = |{\Delta_e}|^2\sum_{\eta, s}^{'} \eta \textrm{sign}(\Delta_s^{\eta})/2 {\Delta_s^{\eta}}^2$,  $|{\Delta_e}|^{^{-1}} = \sum_{\eta, s}^{'}\! |\Delta_s^{\eta}|^{^{-1}}$,  $n$ is the number of closed gaps in a given phase, and $Z_0$ is the vacuum impedance. Similar results hold for $p$-polarized waves by replacing $\sin\!\theta \rightarrow -\sin\!\theta$ and $\cos\!\theta \rightarrow 1/\cos\!\theta$. Note that  topologically trivial phases have the same dependence with the incidence angle $\theta$ and material properties, without distinction between insulating and semimetallic states. Moreover, for $\mathcal{C} = 0$ and small frequencies the first term in  $\Delta_{\rm SHEL}^{\pm ({\rm s})}$ can be neglected. Consequently, the SHEL becomes independent of the optical response of the 2D layer and $\Delta_{\rm SHEL}^{\pm ({\rm s})}$ is maximized for $\theta \simeq \pi/4$. A different landscape takes place when topology enters into play. Unlike the $\mathcal{C} = 0$ case, nontrivial topological states allow for a clear distinction between semimetallic and insulating phases of the system. Indeed, in the AQHI and PS-QHI states the SHEL is a 
 harmonic function of $\theta$ and decays with $\mathcal{C}$. However, at phase transition boundaries a complex interplay between $\theta$ and $\mathcal{C}$ determines the characteristics of the reflected beam.  Also, the fact that one cannot split the contributions of $\mathcal{C}$ and $n$ to $\Delta_{\rm SHEL}^{\pm ({\rm s})}$ in the SPM and SDC states ratifies our conclusion that photonic shifts are sensitive not only to the number of closed gaps but also to the spin and valley numbers of the charge carriers. 
 
Figure \ref{Fig4}a depicts the behavior of the SHEL shifts as a function of the incidence angle at different points of the phase diagram.  Note that for $\mathcal{C} \neq 0$ the incidence angle that maximizes $\Delta_{\rm SHEL}^{\pm ({\rm s})}$ is a function of the material properties 
in each phase.  Figure \ref{Fig4}b shows the dependence of $\Delta^{\pm}_{\rm SHEL}$ with the frequency of the Gaussian beam, clearly illustrating the $1/\omega$ increase of the shifts at long wavelengths (see Eq. (\ref{Expansion_SHEL})). For frequencies $\hbar \omega \gtrsim \lambda_{\rm SO}$ all shifts monotonically decrease, preventing the SHEL to effectively probe distinct topologies in the monolayer. In both Figs. \ref{Fig4}a and  \ref{Fig4}b the approximated results obtained through Eq. (\ref{Expansion_SHEL}) are in excellent agreement with the full numerical calculations given by Eq. (\ref{SHEL_Shift}). In the AQHI phase, however, Eq. (\ref{Expansion_SHEL}) is inaccurate for grazing incidence ($\theta \gtrsim 85^o$)  and frequencies $\hbar\omega \gtrsim 0.5\lambda_{\rm SO}$.  Figures \ref{Fig4}c and \ref{Fig4}d describe how dissipation and temperature affect the SHEL in the graphene family. We note that increasing either $\Gamma$ or $T$ 
results in reduced contrast between electronic phases, although a better distinction between $\Delta^{+}_{\rm SHEL}$ and $\Delta^{-}_{\rm SHEL}$ is achieved. It is also clear that thermal effects have a greater impact on the SHEL
than dissipation. The non-monotonic behavior of $\Delta^{\pm}_{\rm SHEL}$ as we change $T$ is a consequence of the fact that thermal excitations can create electron-hole pairs in the monolayer even if the frequency of the incident  beam does not match any of the mass gaps.
Finally, despite recent progress in the fabrication of free standing stanene \cite{Saxena2016} indicates that our results could soon be tested  in suspended staggered monolayers,  the inclusion of a substrate may be relevant for practical applications. Metallic substrates 
are typically highly reflective in the frequency ranges of interest, thus dominating the SHEL over the topological phase transitions emerging from the 2D semiconductor. On the other hand, a good contrast between the shifts at different electronic phases can be achieved for lossless 
low-refractive index dielectric substrates ($\varepsilon < 1.6$). In Figs. \ref{Fig4}e and \ref{Fig4}f we show the Imbert-Fedorov and relative SHEL shifts
as a function of wavelength $\lambda$ for stanene on top of silicon carbide  \cite{Palik}. Although it is difficult to discern the electronic phases through the SHEL for a $s$-polarized beam (not shown), the effect of topological phase transitions on $\Delta_{\rm IF}$ and $\Delta_{\rm r}$  is clearly appreciated for $p$-polarization and mid-infrared frequencies (in the range shown $0.44 \lesssim \textrm{Re}(\varepsilon) \lesssim 1.52$ and $\textrm{Im}(\varepsilon) < 0.09$). Note that $\Delta_{\rm SHEL}^{\pm} $ can be $ \sim \lambda$ and the position of the peak (zero) of $\Delta_{\rm IF}$  ($\Delta_{\rm r}$) depends on the electronic phase, which is another signature  from the distinct topologies enabled by the graphene family.

In summary,  we have discussed topological phase transitions in the photonic spin Hall effect due to interaction of a Gaussian beam with staggered monolayers of the graphene family. We showed 
that the SHEL 
presents 
signatures of the number of Dirac cones closed and
it depends on the Chern number characterizing the topology of each phase.
Given the reported sensitivities for measuring the SHEL
through weak measurement approaches \cite{Hosten2008}, we conclude that an experimental demonstration of our results
is within current capabilities. 
In spite of the fast experimental progress in the synthesis of staggered 
materials of the graphene family \cite{Gomez2016, Manix2017, Molle2017, PhysRevLett.108.155501,Davila-2014, RIS_0,Saxena2016} suggests that they will be easily accessible in the near future, some of our results can be tested in graphene, as it 
presents topological insulator features under circularly polarized illumination (only QSHI and AQHI phases can be probed in this case)
\cite{Oka2009}.  We envision the effects predicted here
will greatly impact research in 
spinoptics, spintronics, and valleytronics.

\begin{acknowledgments}
The author acknowledges D. Dalvit, L. Woods,  P. Rodriguez-Lopez, and P. Ledwith for discussions and financial support from LANL LDRD and CNLS. 
\end{acknowledgments}


\begin{thebibliography}{99}
\bibitem{comment1} Upon expressing $E_z$, $\Lambda$, $\omega$, $\Gamma$, $T$, and $\mu$ in units of $\lambda_{\rm SO}$ and normalizing the SHEL shifts by $d_0$ all plots become material independent and are valid for any suspended staggered monolayer of the graphene family.
\end{thebibliography}
\end{document}